\documentclass[%
 amsmath,amssymb,
 aps,
]{revtex4-1}

\usepackage{graphicx}
\usepackage{dcolumn}
\usepackage{bm}

\begin{document}

\title{Central charges for Kerr and Kerr-AdS black holes in diverse dimensions}
\author{Jingbo Wang}
\email{ shuijing@mail.bnu.edu.cn}
\affiliation{Institute for Gravitation and Astrophysics, College of Physics and Electronic Engineering, Xinyang Normal University, Xinyang, 464000, P. R. China}
 \date{\today}
\begin{abstract}
In the previous work we give a microscopic explanation of the entropy for the BTZ black hole and four-dimensional Kerr black hole based on the massless scalar field theory on the horizon. An essential input is the central charges of those black holes. In this paper, we calculate the central charges for Kerr black holes and Kerr-AdS black holes in diverse dimensions by rewriting the entropy formula in a suggesting way. Then we also give the statistical explanation for the entropy of those black holes based on the scalar field on the horizon which similar to 4D kerr black hole.
\end{abstract}
\maketitle
\section{Introduction}
The Kerr/CFT correspondence \cite{kerrcft1} is firstly formulated as a duality between extremal Kerr black hole with angular momentum $J$ and a chiral 2D conformal field theory with central charge $c=12J$. Some years later, with the discovery of `hidden conformal symmetry' \cite{kerrcft2} which act on the solution of wave function in near-horizon region, the correspondence extended to Kerr black holes with any values of mass $M$ and $J$. For a good review, see Ref.\cite{kerrcft3}.

The original Kerr/CFT correspondence has been generalized to higher-dimensional Kerr-AdS black holes \cite{kerrads1}. In $D$-dimensional spacetimes higher than four, there are $N=[\frac{D-1}{2}]$ angular momentums, since the black hole can rotate independently in $N$ mutually-orthogonal spatial 2-plane \cite{highd1}. It was found that for each rotation there is a 2D CFT. The central charge is generally different for different rotation.

It was claim by the author that the black holes can be considered as kind of topological insulators \cite{wangti1,wangti2}. Based on this claim, in the previous paper \cite{wangbms4}, the microscopic states of Kerr black hole was given by the quantum scalar field on the horizon. The central charge is assumed to be $c=12 M r_+$ for all Kerr black holes.

In this paper, we calculate the central charges for Kerr black holes and Kerr-AdS black holes in diverse dimensions. Firstly we rewrite the Bekenstein-Hawking entropy formula as sums of Hardy-Ramanujan terms. The central charges can be calculated from this formula. Then from the boundary scalar field we can give the entropy formula a statistical explanation based on counting the underling microscopic states.

The paper is organized as follows. In section II, we analyze the Kerr black holes in diverse dimensions. In section III, the Kerr-AdS black holes are analyzed with the same method. Section IV is the conclusion.
\section{Higher dimensional Kerr black hole}
\subsection{The entropy formula}
The general Kerr metrics in arbitrary higher dimensions were obtained in \cite{highd2,highd3}. The Kerr black holes which include the NUT charges were also obtained in \cite{highd4}. The situation for odd and even spacetime dimension is slightly different, so we consider them separately. Firstly we consider the even case $D=2n$.

The coordinates is chosen to be $(v,r,y_1,\cdots,y_{n-1},\varphi_1,\cdots,\varphi_{n-1})$, where $y_i$ are latitude coordinates and $\varphi_i$ are azimuthal coordinates which have periods $2 \pi$. The explicit form of the metric is unimportant for us, so we just omit it. The horizon is located at $r=r_+$, where $r_+$ is the largest root of the polynomial function
\begin{equation}\label{1}
  X=\prod_{k=1}^{n-1} (r^2+a_k^2)-2 M r.
\end{equation}
The thermodynamics quantities are given by \cite{highd5}
\begin{equation}\label{2}\begin{split}
  E=\frac{(D-2) M V_{D-2}}{8\pi},\quad T_H=\frac{1}{2\pi}[\sum_{i=1}^{n-1} \frac{r_+}{r_+^2+a_i^2}-\frac{1}{2 r_+}], \quad \Omega_i=\frac{a_i }{r_+^2+a_i^2},\\
  S_{BH}=\frac{1}{4}V_{D-2}\prod_{i=1}^{n-1} (r_+^2+a_i^2)=\frac{1}{2}V_{D-2} M r_+,\quad J_i=\frac{a_i V_{D-2} M}{4\pi},
\end{split}\end{equation}
where $a_i$ are rotation parameters, $V_{D-2}$ are volume of the unit $(D-2)$-sphere, and $\Omega_i$ are angular velocities measured in the asymptotically non-rotating frame. They satisfy the so-called Quantum Statistical Relation \cite{highd5}, that is,
\begin{equation}\label{2a}
  E-T S-\sum_{k}\Omega_k J_k=T I_{D-2}\equiv \Phi=\frac{M V_{D-2}}{8\pi},
\end{equation}
where $I_{D-2}$ is the action of the Kerr black hole, and $\Phi$ the thermodynamic potential.

There are $n-1$ independent rotations in the $n-1$ orthogonal spatial $2$-planes.  The following equality is important
\begin{equation}\label{3}
 \gamma_i \Phi r_+\pm J_i=\frac{(r_+^2+a_i^2)V_{D-2}}{8\pi r_+} M\pm J_i=\frac{V_{D-2} M}{8\pi r_+}(r_+\pm a_i)^2\geq 0,\quad \gamma_i\equiv\frac{r_+^2+a_i^2}{r_+^2}.
\end{equation}
Just like that in 4D Kerr black hole case \cite{wangbms4}, for each rotation one can rewrite the entropy as follows
\begin{equation}\label{4}
  S_{BH}=2\pi \sqrt{\frac{N_{2 k-1}}{6}}+2\pi \sqrt{\frac{N_{2 k}}{6}},
\end{equation}
where
\begin{equation}\label{5}
  N_{2 k-1}=\frac{c_k}{4}(\frac{(r_+^2+a_k^2)V_{D-2}}{8\pi r_+} M+ J_k)=\frac{c_k}{4}(\gamma_i \Phi r_++ J_k),\quad N_{2 k}=\frac{c_k}{4}(\frac{(r_+^2+a_k^2)V_{D-2}}{8\pi r_+} M- J_k)=\frac{c_k}{4}(\gamma_i \Phi r_+- J_k).
\end{equation}
From the above equation, we will get the central charge
 \begin{equation}\label{6}
   c_k=c=\frac{6 M r_+ V_{D-2}}{2 \pi }=\frac{6 S_{BH}}{\pi}
 \end{equation}
for all $k$.

For Schwarzschild black hole, the central charge is
 \begin{equation}\label{6c}
   c=\frac{3 r_+^{D-2} V_{D-2}}{2 \pi }.
 \end{equation}
\subsection{The microscopic explanation}
 Next we want to give the formula (\ref{4}) a microscopic explanation. The boundary degrees of freedom of the Kerr black hole can be described by a BF theory, which can be cast into a massless scalar field theory \cite{wangbms4}
 \begin{equation}\label{6b}\begin{split}
    S'=\frac{m_0}{2}\int_{\Delta}d^{D-1} x\sqrt{-g}g^{\mu\nu}\partial_\mu \phi \partial_\nu \phi.
\end{split}\end{equation}

For each rotation $k$ the effect metric on the horizon is assumed to be \cite{highd4}
 \begin{equation}\label{6a}
   \tilde{ds}_k^2=-dv_k'^2+r_+^2 dV_{D-2}^2=-dv_k'^2+r_+^2(\sum_{j=1}^{n-1} g_j dy_j^2+\sum_{i=1}^{n-1} \tilde{g}_i d\varphi_i^2),
 \end{equation}
 where $v_k'=\frac{v}{\gamma_k}$, $g_j,\tilde{g}_i$ are complicated functions of the $(a_i,y_j)$.

 The massless scalar field $\phi(v_k',y_i,\varphi_i)$ has the mode expanding
 \begin{equation}\label{7}
  \phi(v_k',y_i,\varphi_i)=\phi_0+p_v v'+\sum_i p_{y_i} f(y_i)+\sum_j p_{\varphi_j} \varphi_i+\sqrt{\frac{1}{m_0 A}}\sum_{l> 0}\sum_{m_i}\sqrt{\frac{1}{2\omega_l}}[a_{l,m_i} e^{-i \omega_l v'+i m_i \varphi_i}F(y_i)+a^*_{l,m_i} e^{i \omega_l v'-i m_i \varphi_i}F(y_i)^*],
 \end{equation}
 where $e^{i m_i \varphi_i}F(y_i)$ and  $e^{i m_i \varphi_i}f(y_i)$ satisfy the Laplace equation on $S_{D-2}$ with
 \begin{equation}\label{8}
   \Delta_{S_{D-2}} [e^{i m_i \varphi_i}F(y_i)]=-l(l+D-3) [e^{i m_i \varphi_i}F(y_i)],\quad \Delta_{S_{D-2}} [e^{i m_i \varphi_i}f(y_i)]=0.
 \end{equation}

 The Hamiltonian for the free scalar field can be written as
\begin{equation}\label{9}\begin{split}
  H_{free}=\frac{m_0}{2}\int_{\Sigma}d^{D-2} x\sqrt{-g}(\partial_0 \phi \partial_0 \phi+g^{ij}\partial_i \phi \partial_j \phi)\\
  = H_0+\sum_{l> 0}\sum_{m_i} \frac{\sqrt{l(l+D-3)}}{r_+}a^+_{l,m_i} a_{l,m_i}\\
  \end{split}\end{equation}
where $H_0$ are zero mode contribution and we omit the zero-point energy. In high dimensional spacetime $D\geq 4$, general relativity have local degrees of freedom. So it is natural to consider scalar field with interaction
\begin{equation}\label{9a}
  H_{full}=H_{free}+H_{int}.
\end{equation}

We can also define the angular momentum operator for each rotation
\begin{equation}\label{10}\begin{split}
 J_i=m_0 \int_{\Sigma}d^{D-2} x\sqrt{-g} (\partial_0 \phi \partial_i \phi)\\
 =J_{i0}+\sum_{l\> 0}\sum_{I}m_i^I a^+_{l,m_i^I} a_{l,m_i^I},
\end{split}\end{equation}
where $J_{i0}$ are zero mode contribution.

 The scalar field $\phi(v_k',y_i,\varphi_i)$ can be considered as collectives of harmonic oscillators, and a general quantum state can be represented as $|p_v,p_{y_i},p_{\varphi_i};\{n_{l,m_i}\}>$ where $p_v,p_{y_i},p_{\varphi_i}$ are zero mode part, and $\{n_{l,m_i}\}$ are oscillator part. Here we make an important assumption: the zero mode part and the oscillator part have same energy and angular momentum. The Kerr black hole with parameters $(M,J)$ corresponds to the zero mode part, thus
\begin{equation}\label{57}
  <p_v,p_{y_i},p_{\varphi_i};\{0\}|\hat{J}_i|p_v,p_{y_i},p_{\varphi_i};\{0\}>=\frac{J_i}{2},\quad <p_v,p_{y_i},p_{\varphi_i};\{0\}|\hat{H}^i_{full}|p_v,p_{y_i},p_{\varphi_i};\{0\}>=\frac{\gamma_i \Phi}{2}.
\end{equation}

To move on, we need to know the spectrum of the full hamiltonian $\hat{H}^i_{full}$. From the equality (\ref{3}) and the constraints (\ref{11}), one can conjecture that the spectrum of the full Hamiltonian is
\begin{equation}\label{11a}\begin{split}
  H^i_{full}= H_0+\sum_{l> 0}\sum_{I} \frac{|m_i^I|}{r_+}a^+_{l,m_i^I} a_{l,m_i^I}\\
  =H^i_0+\sum_I \frac{|m_i^I|}{r_+}\hat{n}_{m_i^I}, \quad m^I \neq 0,
  \end{split}\end{equation}
  where $\hat{n}_{m_i^I}=\sum_{l>0} \hat{n}_{l,m_i^I}, \hat{n}_{l,m_i^I}=\hat{a}^+_{l,m_i^I} \hat{a}_{l,m_i^I}$ are number operators. The angular momentum operator can also be rewritten as
\begin{equation}\label{55c}\begin{split}
    \hat{J}_i=\hat{J}_0+\sum_I m_i^I \hat{n}_{m_i^I}, \quad m^I \neq 0.
 \end{split}\end{equation}

The microstates of Kerr black hole can be represented by $|0,0;\{n_{m_i^I}\}>$. Similar to the 4D kerr black hole, for each rotation, we require
\begin{equation}\label{11}
 \frac{1}{c_i} <0,0;\{n_{m_i^I}\}|\hat{J}_i|0,0;\{n_{m_i^I}\}>=\frac{J_i}{2}, (i=1,\cdots,n-1),\quad \frac{1}{c_i}<0,0;\{n_{m_i^I}\}|\hat{H}^i_{full}|0,0;\{n_{m_i^I}\}>=\frac{\gamma_i \Phi}{2},
\end{equation}
where $c_i$ is the central charge and $\hat{H}^i_{full}$ the Hamiltonian associated with the rotation $i$. Different sequence $\{n_{m_i^I}\}$ corresponds to different microstate of the Kerr black hole with same $(M,J_i)$.

 With quantum number, the constraints (\ref{11}) can be written as
\begin{equation}\label{12}
  \sum_I |m_i^I| n_{m_i^I}=c_i \frac{\gamma_i \Phi r_+}{2},\quad \sum_I m_i^I n_{m_i^I}=c_i \frac{J_i}{2},  \forall 1\leq i \leq n-1.
\end{equation}
For each $i$, we calculate the number of different sequence $\{n_i^I\}$.

The number can be given by the Hardy-Ramanujan formula,
\begin{equation}\label{14}
  N(J_i) \simeq \exp(2\pi \sqrt{\frac{N_{2i-1}}{6}}+2\pi \sqrt{\frac{N_{2i}}{6}}),
\end{equation}
with $N_{2i-1}$ and $N_{2i}$ are given by (\ref{5}). Thus, the microscopic states of the massless scalar field can explain the entropy of the Kerr black holes.

Next we consider the odd case $D=2n-1$. The horizon is given by the condition
\begin{equation}\label{15}
  X=\prod_{k=1}^{n-1} (r^2+a_k^2)-2 M r^2=0.
\end{equation}
The thermodynamics quantities are given by
\begin{equation}\label{16}\begin{split}
  E=\frac{(D-2) M V_{D-2}}{8\pi},\quad T_H=\frac{1}{2\pi}[\sum_{i=1}^{n-1} \frac{r_+}{r_+^2+a_i^2}-\frac{1}{r_+}], \quad J_i=\frac{a_i V_{D-2} M}{4\pi},\\
  S_{BH}=\frac{1}{4 r_+}V_{D-2}\prod_{i=1}^{n-1} (r_+^2+a_i^2)=\frac{1}{2}V_{D-2} M r_+,\quad \Omega_i=\frac{a_i }{r_+^2+a_i^2},
\end{split}\end{equation}
where $a_i$ are rotation parameters. The left are the same for $D=2n$ case. The central charge is also given by the formula (\ref{6}).
\section{Kerr-AdS black holes}
The generalization to Kerr-AdS black holes in straightforward.

For $D=2n$. The coordinates is also chosen to be $(v,r,y_1,\cdots,y_{n-1},\varphi_1,\cdots,\varphi_{n-1})$. The horizon is located at $r=r_+$, where $r_+$ is the largest root of the polynomial function
\begin{equation}\label{17}
  X=(1+\frac{r^2}{L^2})\prod_{k=1}^{n-1} (r^2+a_k^2)-2 M r.
\end{equation}
The thermodynamics quantities are given by
\begin{equation}\label{18}\begin{split}
  E=\frac{M V_{D-2}}{4\pi(\prod_j \Xi_j)}\sum_{i=1}^{n-1}\frac{1}{\Xi_i},\quad T_H=\frac{1}{2\pi}[(1+\frac{r_+^2}{L^2})\sum_{i=1}^{n-1} \frac{r_+}{r_+^2+a_i^2}-\frac{1-\frac{r_+^2}{L^2}}{2 r_+}], \quad J_i=\frac{a_i V_{D-2} M}{4\pi \Xi_i (\prod_j \Xi_j)},\\
  S_{BH}=\frac{1}{4}V_{D-2}\prod_{i=1}^{n-1}\frac{r_+^2+a_i^2}{\Xi_i}=\frac{1}{2}V_{D-2} M r_+ \frac{1}{(1+\frac{r_+^2}{L^2}) (\prod_j \Xi_j)},\quad \Omega_i=\frac{a_i(1+\frac{r_+^2}{L^2}) }{r_+^2+a_i^2},
\end{split}\end{equation}
where $\Xi_i=1-\frac{a_i^2}{L^2}$. The thermodynamical potential is given by
\begin{equation}\label{2a}
 \Phi=E-T S-\sum_{k}\Omega_k J_k=\frac{M V_{D-2}}{8\pi (\prod_j \Xi_j)}\frac{1-\frac{r_+^2}{L^2}}{1+\frac{r_+^2}{L^2}}.
\end{equation}

Notice the equality
\begin{equation}\label{19}
 \gamma_i \Phi r_+\pm J_i=\frac{(r_+^2+a_i^2)V_{D-2}}{8\pi r_+ \Xi_i (\prod_j \Xi_j)} M\pm J_i=\frac{V_{D-2} M}{8\pi r_+ \Xi_i (\prod_j \Xi_j)}(r_+\pm a_i)^2\geq 0,\quad  \gamma_i\equiv\frac{r_+^2+a_i^2}{r_+^2 \Xi_i}\frac{1+\frac{r_+^2}{L^2}}{1-\frac{r_+^2}{L^2}}.
\end{equation}
One can rewrite the entropy as follows
\begin{equation}\label{20}
  S_{BH}=2\pi \sqrt{\frac{N_{2 k-1}}{6}}+2\pi \sqrt{\frac{N_{2 k}}{6}},
\end{equation}
where
\begin{equation}\label{21}\begin{split}
  N_{2 k-1}=\frac{c_k}{4} (\frac{(r_+^2+a_k^2)V_{D-2}}{8\pi r_+\Xi_i (\prod_j \Xi_j)} M+ J_k)=\frac{c_k}{4} (\gamma_i \Phi r_++ J_k),\\
  N_{2 k}=\frac{c_k}{4} (\frac{(r_+^2+a_k^2)V_{D-2}}{8\pi r_+\Xi_i (\prod_j \Xi_j)} M- J_k)=\frac{c_k}{4} (\gamma_i \Phi r_+- J_k).
\end{split}\end{equation}
It gives the central charge for rotation $k$
 \begin{equation}\label{22}
   c_k=\frac{6 M r_+ V_{D-2}}{2 \pi }\frac{\Xi_k}{(1+\frac{r_+^2}{L^2})^2 (\prod_j \Xi_j)}=\frac{6 S_{BH}}{\pi }\frac{1-\frac{a_k^2}{L^2}}{1+\frac{r_+^2}{L^2}}.
 \end{equation}

 \section{Conclusion}
 In this paper, we study the central charges of Kerr black holes and Kerr-AdS black holes in diverse dimensions. After rewriting the entropy as a suggestive form, we give the central charges (\ref{6}) and (\ref{22}) for Kerr and Kerr-AdS black hole respectively. We also give a microscopic explanation of this entropy based on the quantum states of the interact scalar field on the horizon. One important conjecture is about the spectrum of the full Hamiltonian for each rotation (\ref{11a}). This special form may give some clues for microscopic dynamics of black hole.
\acknowledgments
 This work is supported by Nanhu Scholars Program for Young Scholars of XYNU.


\end{document}